\documentclass[%
 aip,
 amsmath,amssymb,
 reprint,%
]{revtex4-1}

\usepackage{graphicx}% Include figure files
\usepackage{dcolumn}% Align table columns on decimal point
\usepackage{bm}% bold math
\usepackage[utf8]{inputenc}
\usepackage[T1]{fontenc}
\usepackage{mathptmx}
\usepackage{etoolbox}
\usepackage{siunitx}
\usepackage{layouts}
\usepackage{comment}
\usepackage{import}
\usepackage{chemformula}
\usepackage{ifthen}

\newcommand{\mode}{show_revs} %change to sth else to remove the highlighting of changes
\ifthenelse{\equal{\mode}{show_revs}}
  {% True case
  }
  {% false case
  }

\makeatletter
\def\@email#1#2{%
 \endgroup
 \patchcmd{\titleblock@produce}
  {\frontmatter@RRAPformat}
  {\frontmatter@RRAPformat{\produce@RRAP{*#1\href{mailto:#2}{#2}}}\frontmatter@RRAPformat}
  {}{}
}%
\makeatother
\begin{document}

\preprint{AIP/123-QED}
%TC:ignore
\title[Noise reduction by bias cooling]{Noise reduction by bias cooling in gated \ch{Si}/\ch{Si_x Ge_{1-x}} quantum dots}
% Force line breaks with \\
\author{Julian Ferrero}
 \affiliation{ 
Physikalisches Institut, Karlsruhe Institute of Technology, Karlsruhe, Germany
}%
\author{Thomas Koch}
 \affiliation{ 
Physikalisches Institut, Karlsruhe Institute of Technology, Karlsruhe, Germany
}%
\author{Sonja Vogel}%
 \affiliation{ 
Physikalisches Institut, Karlsruhe Institute of Technology, Karlsruhe, Germany
}%
\author{Daniel Schroller}%
 \affiliation{ 
Physikalisches Institut, Karlsruhe Institute of Technology, Karlsruhe, Germany
}%
\author{Viktor Adam}%
 \affiliation{ 
Physikalisches Institut, Karlsruhe Institute of Technology, Karlsruhe, Germany
}
 \affiliation{ 
Institute for Quantum Materials and Technologies, Karlsruhe Institute of Technology, Karlsruhe, Germany
}%
\author{Ran Xue}%
 \affiliation{ 
JARA-FIT Institute for Quantum Information, Forschungszentrum Jülich GmbH and RWTH Aachen University, Aachen, Germany%\\This line break forced with \textbackslash\textbackslash
}%
\author{Inga Seidler}%
 \affiliation{ 
JARA-FIT Institute for Quantum Information, Forschungszentrum Jülich GmbH and RWTH Aachen University, Aachen, Germany%\\This line break forced with \textbackslash\textbackslash
}%
\author{Lars R. Schreiber}%
 \affiliation{ 
JARA-FIT Institute for Quantum Information, Forschungszentrum Jülich GmbH and RWTH Aachen University, Aachen, Germany%\\This line break forced with \textbackslash\textbackslash
}%
\author{Hendrik Bluhm}%
 \affiliation{ 
JARA-FIT Institute for Quantum Information, Forschungszentrum Jülich GmbH and RWTH Aachen University, Aachen, Germany%\\This line break forced with \textbackslash\textbackslash
}%
\author{Wolfgang Wernsdorfer}
 \email{wolfgang.wernsdorfer@kit.edu}
 \affiliation{ 
Physikalisches Institut, Karlsruhe Institute of Technology, Karlsruhe, Germany
}
 \affiliation{ 
Institute for Quantum Materials and Technologies, Karlsruhe Institute of Technology, Karlsruhe, Germany
}%

\date{\today}% It is always \today, today,
\begin{abstract}
Silicon-Germanium heterostructures are a promising quantum circuit platform, but crucial aspects as the long-term charge dynamics and cooldown-to-cooldown variations are still widely unexplored quantitatively. In this letter we present the results of an extensive bias cooling study performed on gated silicon-germanium quantum dots with an \ch{Al2O3}-dielectric. Over 80 cooldowns were performed in the course of our investigations. The performance of the devices is assessed by low-frequency charge noise measurements in the band of \SI{200}{\micro\hertz} to \SI{10}{\milli\hertz}. We measure the total noise power as a function of the applied voltage during cooldown in four different devices and find a minimum in noise at \SI{0.7}{\volt} bias cooling voltage for all observed samples. We manage to decrease the total noise power median by a factor of 6 and compute a reduced tunneling current density using Schrödinger-Poisson simulations. Furthermore, we show the variation in noise from the same device in the course of eleven different cooldowns performed under the nominally same conditions. 
\end{abstract}
%TC:endignore
\maketitle
Electron spins in silicon-germanium represent a promising implementation of the solid state quantum computer. Single and two-qubit gates have been reported to reach gate fidelities above the error correction threshold \cite{Yoneda2018, Noiri2022, Xue2024}. Recently, quantum error correction was performed on a three qubit device \cite{Takeda2022} and intermediate range coupling opened new prospects regarding a scaled up quantum processor\cite{Seidler2022, Langrock2023, Xue2024, Kuenne2023}. Since quantum-dot based qubits need to be tuned and re-tuned, the long term stability of the qubit working point has gained research interest \cite{Kranz2020, Struck2020, Connors2022}. Furthermore it has been shown that the charge noise level in \ch{SiGe} devices is strongly dependent on the applied global field \cite{Takeda2013}. Bias cooling is a readily accessible method which only relies on equipment already present in the typical semiconductor quantum dot setup. To perform bias cooling, the same bias-cooling-voltage $\mathrm{V_{BC}}$ is applied to all gates of the sample during cooldown from room-temperature to \si{\milli\kelvin} temperatures. This causes charges to be trapped in localized defects which interact with the global electric field, thereby changing the stability of gate defined quantum dots. In gallium arsenide, bias cooling has been proven to reduce switching noise and therefore improving the sample stability \cite{PioroLadriere2005}, by filling DX centers in the dopant layer and therefore reducing the effective leakage rate of electrons from the gate layers into the two-dimensional-electron-gas (2DEG). In silicon-germanium, no such centers are present and bias cooling has not been quantitatively investigated yet.

We present the results of a systematic bias-cooling study, quantitatively investigating the noise power in the voltage range in between \SI{-1}{\volt} and \SI{1}{\volt} for $\mathrm{V_{BC}}$. More then 80 cooldowns were performed, on four different samples. Depending on the applied gate voltage during cooldown, we reduce the total noise power in the frequency band of \SI{200}{\micro\hertz} to \SI{10}{\milli\hertz} by a factor of 6, in comparison to the zero volt case.
\begin{figure*}[h!]
    \includegraphics{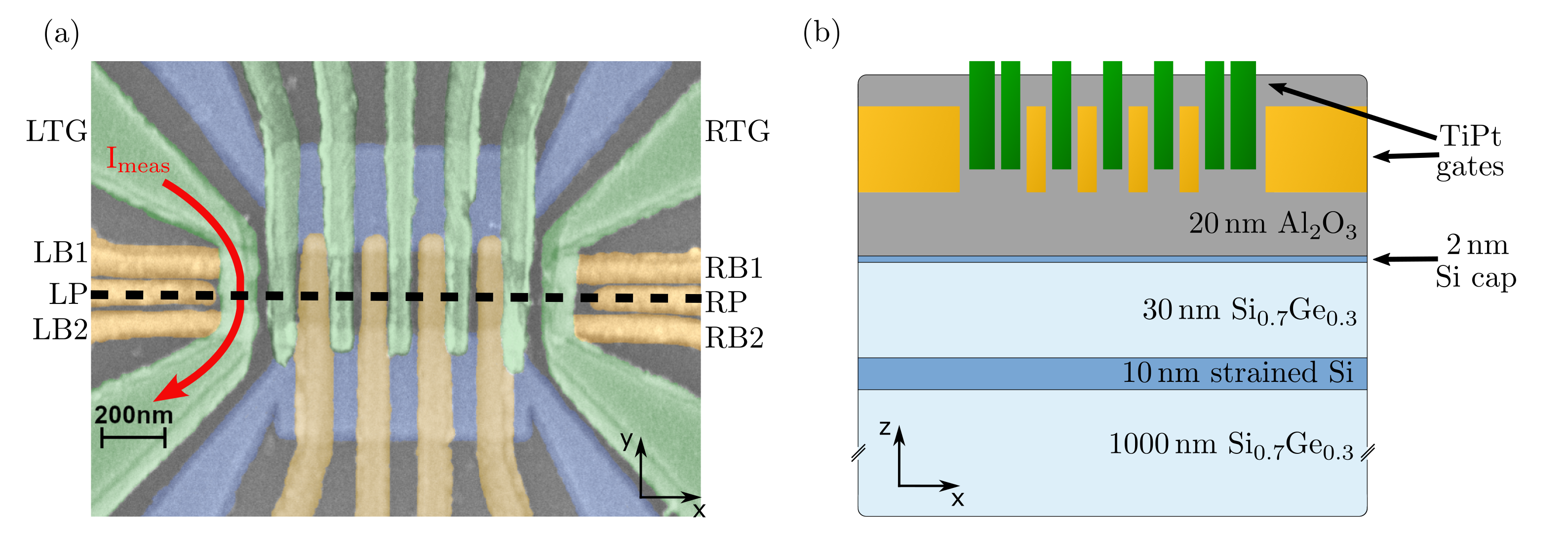}% Here is how to import EPS art
    \caption{\label{fig:sample_details} (a) False colored scanning electron micrograph. Bottommost gate layer in blue, topmost layer in green. The path of the current $\mathrm{I_{meas}}$ used in the  measurements is denoted in red. (b) Schematic cross-section with layer thicknesses to scale. The 2DEG is defined in the \SI{10}{\nano\meter} strained Si layer. The SiGe spacer layer is protected from oxidation by a \SI{2}{\nano\meter} silicon cap layer. The dashed line in (a) represents the corresponding position of the schematic cross-section in (b).}
\end{figure*}
We use four gated Si/SiGe devices nominally equal to the device used in \cite{Seidler2022, Struck2024} (see Fig. \ref{fig:sample_details} (a)). It contains two single-electron transistors (SETs) operating as proximal charge detectors and nine central finger gates as well as two confinement gates.

The gatestack is fabricated upon a Si/SiGe heterostructure, which consists of a \SI{1000}{\nano\meter} \ch{Si_{0.7}Ge_{0.3}} graded buffer layer on a Si substrate,  a \SI{10}{\nano\meter} $\mathrm{Si}$ channel, followed by a \SI{30}{\nano\meter} \ch{Si_{0.7}Ge_{0.3}} spacer layer and capped by \SI{2}{\nano\meter} naturally oxidized Si (see Fig.\ref{fig:sample_details} (b)). The heterostructure confines electrons in the growth direction, effectively forming a 2DEG in the silicon channel, when accumulated. Ohmic contacts to the quantum well layer are realized by implantation of phosphorus and activated by rapid thermal processing at 700 $^{\circ}\mathrm{C}$ for \SI{15}{\second}. The gate stack is an overlay of three metal layers consisting of \SI{15}{\nano\meter}, \SI{22}{\nano\meter} and \SI{29}{\nano\meter} Pt on top of a \SI{5}{\nano\meter} Ti adhesion layer. The metal layers are  electrically insulated from the substrate and from each other by \SI{10}{\nano\meter} atomic layer deposited \ch{Al2O3}.

To characterize the performance under different bias cooling conditions, an efficient thermal cycling mechanism is needed. A heater consisting of constantan wire was installed on the mixing chamber of the used dilution refrigerator. The sample is heated locally to \SI{300}{\kelvin}, while the 4\,K stage of the fridge is kept below \SI{10}{\kelvin}. This is achieved by thermally insulating the mixing chamber plate by stopping the circulation of the mixture while a high flow of liquid helium is supplied to the 4K stage. An automated thermal cycle from \SI{30}{\milli\kelvin} to \SI{300}{\kelvin} and back to \SI{30}{\milli\kelvin} takes \SI{3,5}{\hour}.

\begin{figure} 
    \includegraphics{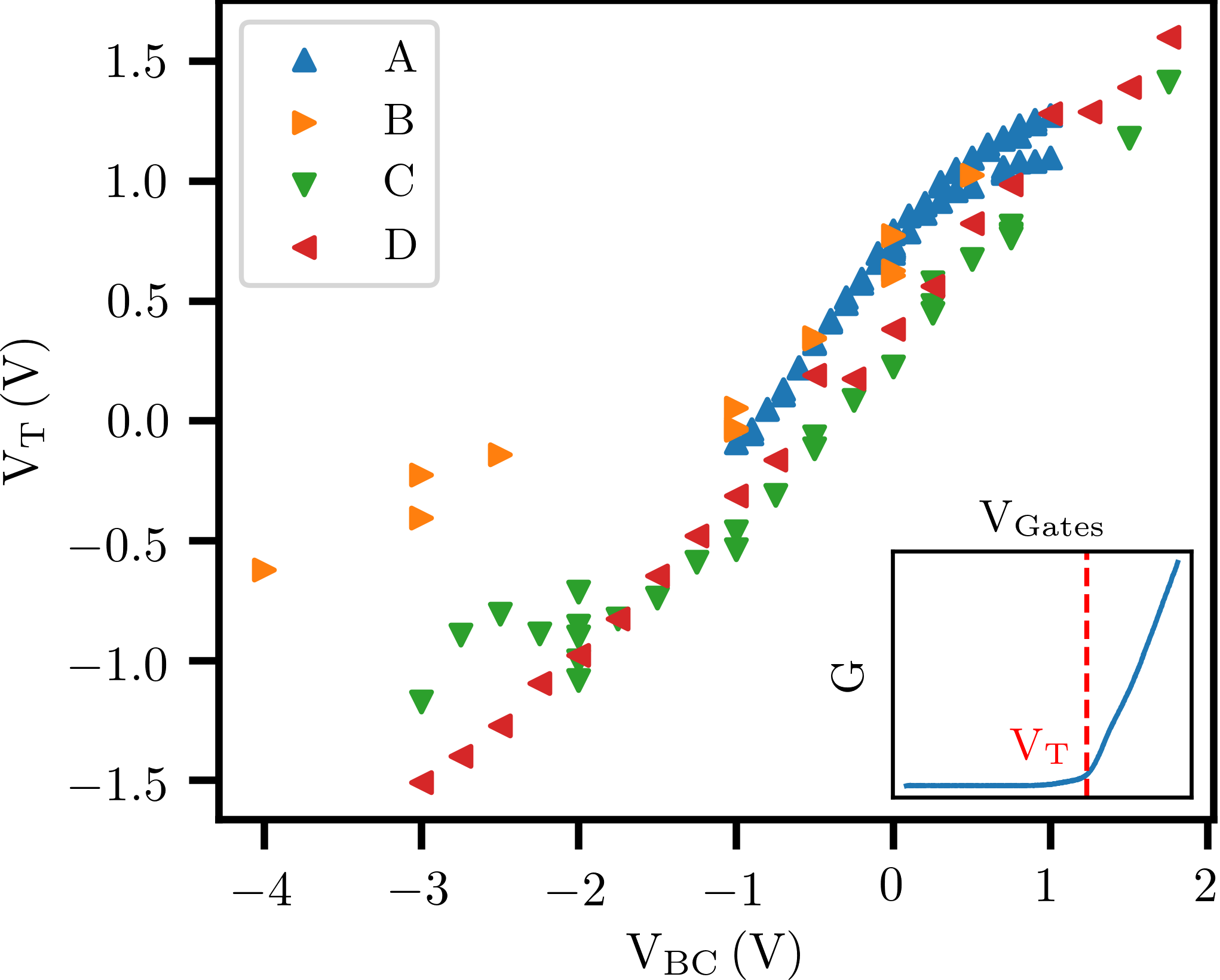}
    \caption{\label{fig:AccV_BCV} Turn-on voltage of the device over applied bias voltage during cooldown for four nominally equal devices. In between \SI{-1}{\volt} and \SI{1}{\volt} all samples behave linearly showing similar slopes. Below \SI{-1}{\volt} the behavior differs. This might be due to different defect densities in the heterostructure. The inset shows an exemplary accumulation curve. The red-dashed line indicates the turn-on voltage, which is defined by the channel conductivity $\mathrm{G}$ crossing a value of \SI{0.4}{\micro\siemens}.}
\end{figure}

\begin{figure*} 
    \includegraphics{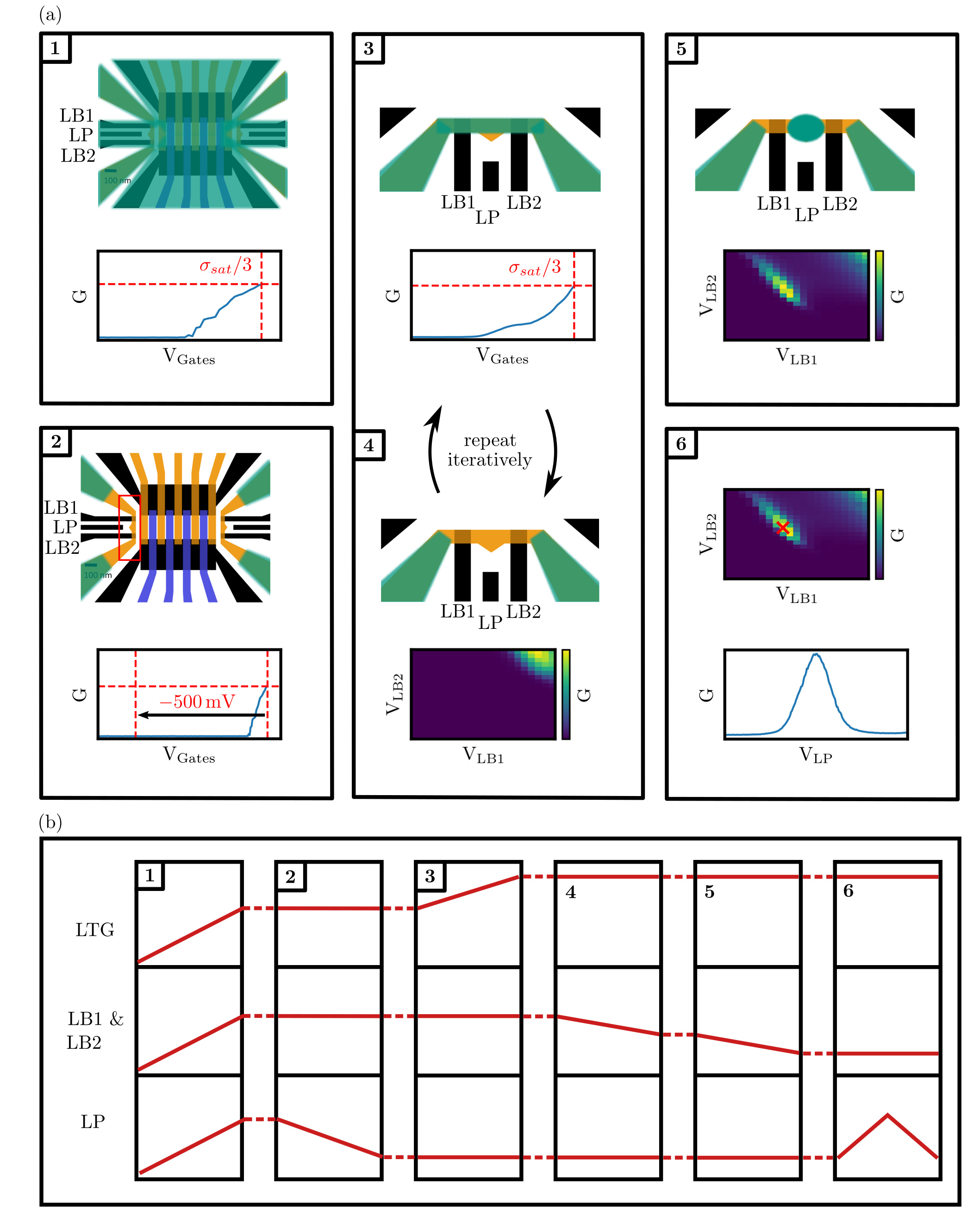}
    \caption{\label{fig:tuning} (a) Tuning workflow which was performed after each cooldown. The red square in step 2 denotes where the cutout shown in steps 3 to 5 lies within sample. The sample is accumulated to the reference conductance of $\mathrm{G_{sat}/3}$ (step 1). The voltages applied to all but the SET-gates are reduced by \SI{500}{\milli\volt} to confine the 2DEG (step 2). Using the SET gates a closed conducting channel is formed (step 3). A 100x\SI{100}{\milli\volt\squared} sweep is performed with both barrier gates (step 4). Step 3 and 4 are repeated iteratively until a Coulomb oscillation is observed (step 5). The barrier gates are tuned to the first Coulomb peak (step 6) and a plunger trace is recorded. (b) Simplified chronogram of the voltages applied to the SET-gates. Numbers 1-6 correspond to the tuning steps in panel (a).}
\end{figure*}

We monitor the turn-on voltage $\mathrm{V_T}$ after reaching base temperature. To do this, we accumulate the electron gas using all gates, effectively forming a conductive sheet underneath the gate structure. This way, we minimize the effect of singular defects on the conducting channel. We measure the conductance $\mathrm{G}$ in between the ohmic contacts of the left SET. Once it reaches one third of the saturation value ($\mathrm{G_{sat}/3}$), which was measured in a separate cooldown, the accumulation voltage is not increased further. The first apparent effect of bias cooling is that the accumulation voltage of the device shifts nearly linearly with the applied bias cooling voltage (Fig. \ref{fig:AccV_BCV}). Charge carriers are trapped in between the quantum well and the metal gates. We suspect electrons getting caught at the Si-\ch{Al2O3} interface \cite{Hoex2008, Gielis2008}. This interface is known to form an \ch{SiO_x} layer, in which silicon atoms are partially replaced by aluminium atoms \cite{Hiller2019, Hiller2021}. These form acceptor states, which trap electrons. By applying the bias cooling voltage $\mathrm{V_{BC}}$ during cooldown, we change the electrochemical potential of the defects, allowing thermally excited electrons to change the population of the defect states. Applying positive voltages attracts additional electrons into acceptor-states, increasing the turn-on voltage. Negative biases reduce the population of the \ch{Al}-vacancies below its equilibrium, creating an excess of positively charged defects, decreasing the accumulation voltage.
Next, we characterize the noise power for each bias voltage. To investigate the charge noise we form a single electron transistor in between barrier gates (e.g. LB1 and LB2). The gate layout of the used device is shown in Fig. \ref{fig:sample_details} (b). The challenge hereby is tuning the SETs to comparable working points, although the sample provides a different electrostatic environment in each cooldown. Noise seen by SET devices depends on their working point \cite{Elsayed2022}. Therefore it is essential to repeat a fixed measurement protocol. Our protocol consists of six steps summarized in Fig. \ref{fig:tuning} (a) and (b). After accumulating the sample to the reference conductance of $\mathrm{G_{sat}/3}$ (step 1), we define the accumulation voltage $V_{\mathrm{ref}}$ which is used as a starting point for the device tuning. With all gates at $\mathrm{V_{ref}}$ the device is in a state where a 2DEG is accumulated in the silicon quantum well below every metal gate. In step 2, we lower the voltages of all gates except the topgates and barriers of the SETs by \SI{500}{\milli\volt} to deplete the sample and confine the 2DEG only below the top gates. In the first iteration of step 3, the top gate and barrier voltages are increased until a third of the saturation conductance is reached (not shown in Fig 3(b)). For every subsequent iteration of step 3, only the top gate is raised until accumulation (Fig 3 (b)). In step 3, the topgate voltage (and in the first iteration also the barrier voltages, not shown in the voltage chronogram in Fig. \ref{fig:tuning} (b)) are increased until a third of the saturation conductance is reached in the measured channel. In step 4 the 'cutoff' voltage of the barrier gates is determined by lowering the voltages of the barrier gates to the point where the conducting channels are fully confined and the measured current is cut off. Slightly above this cutoff voltage, we perform a 100x\SI{100}{\milli\volt\squared} sweep downwards with both barrier gates. We record the current through the channel to see whether Coulomb oscillations are present. If not we now lower the barrier gates by another \SI{100}{\milli\volt} and re-accumulate the conducting channel with the top gate again up to a conductance of $\mathrm{G_{sat}/3}$. By repeating steps 3 and 4 iteratively we find the lowest (granulated by the resolution of our voltage steps) topgate voltage that allows formation of quantum dots, which are identified by a 2D barrier sweep exhibiting the Coulomb oscillations (step 5). The barrier gates are tuned to the first Coulomb peak (step 6) and a plunger trace is recorded. If the recorded trace shows a secant-like Coulomb peak, the SET has been formed and the tuning routine came to a successful conclusion.

\begin{figure}[h]
    \includegraphics{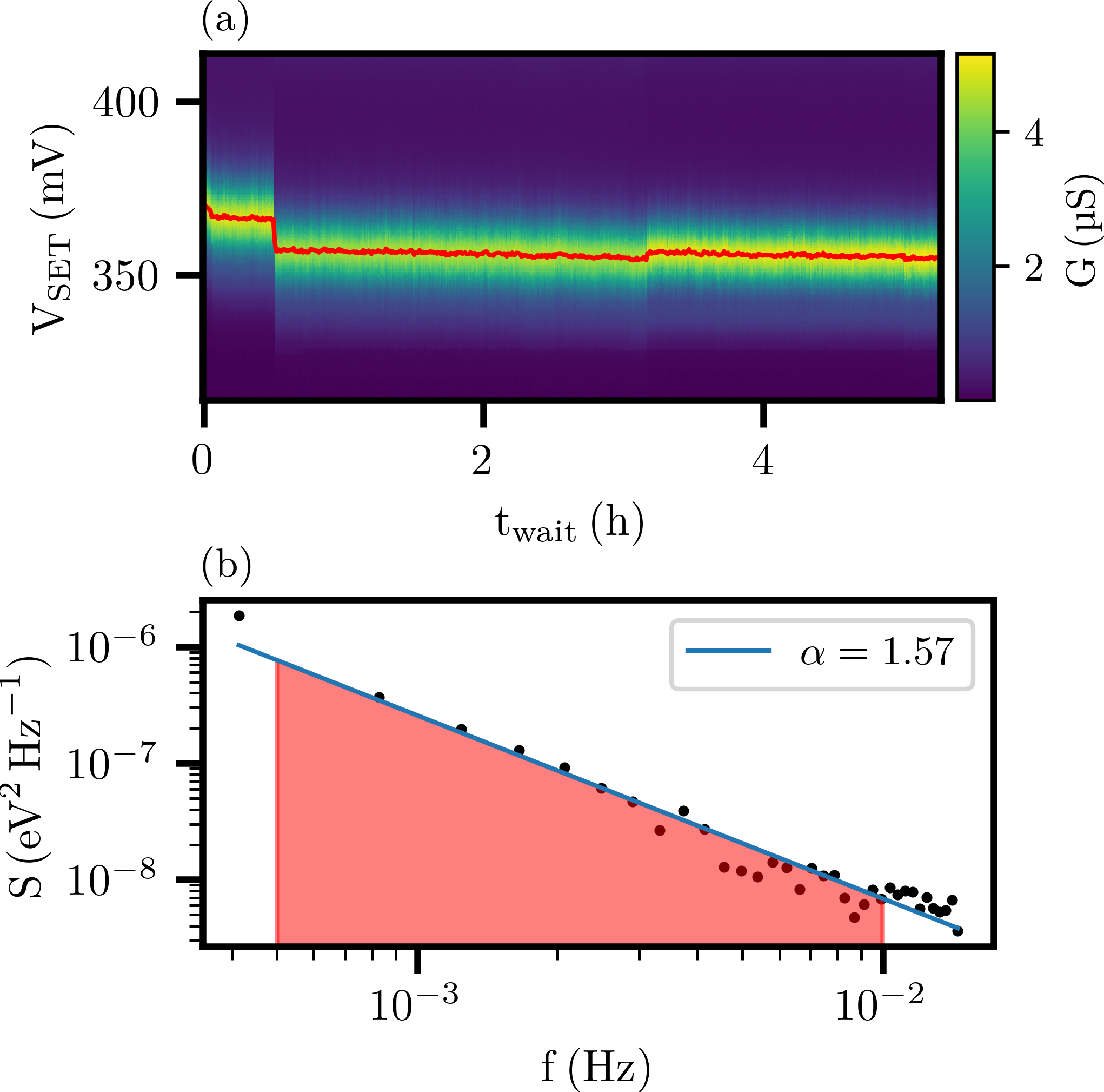}% Here is how to import EPS art
    \caption{\label{fig:metrics_xplain}(a) Peaktracking measurement, performed on sample A with a $\mathrm{V_{BC}}$ of \SI{0}{\volt}. The colorplot consists of Coulomb peak traces recorded back to back. The red line indicates the position of the peak maxima. b) Noise power spectral density, fitted with a $\beta/f^\alpha$ power spectrum. To calculate the total noise power, the spectrum is integrated over the highlighted frequency range. The conversion of $\mathrm{V_{SET}}$ to the chemical potential of the SET is done by a static leverarm of 0.039\,eV/V.}   
\end{figure}

With the quantum dot formed, we characterize its noise in the band of \SI{200}{\micro\hertz} to \SI{10}{\milli\hertz} using peaktracking\cite{Kranz2020}. The lower bound on the measured frequencies is set by the total length of our peaktracking measurements (\SI{5}{\hour}) and the upper bound is set by the rate at which the plunger gate can be swept to record the position of the peak in gatespace. The number of points recorded per plunger sweep varies depending on the expected drift of the sample. Therefore, a trace can contain 150 to 250 individual data points and be measured in \SI{30}{\second} to \SI{70}{\second} depending on the number and sampling rate of the points. We continuously sweep the plunger and record the conductivity of the SET device (Fig. \ref{fig:metrics_xplain} (a)).
\begin{figure*}[h!]
    \includegraphics{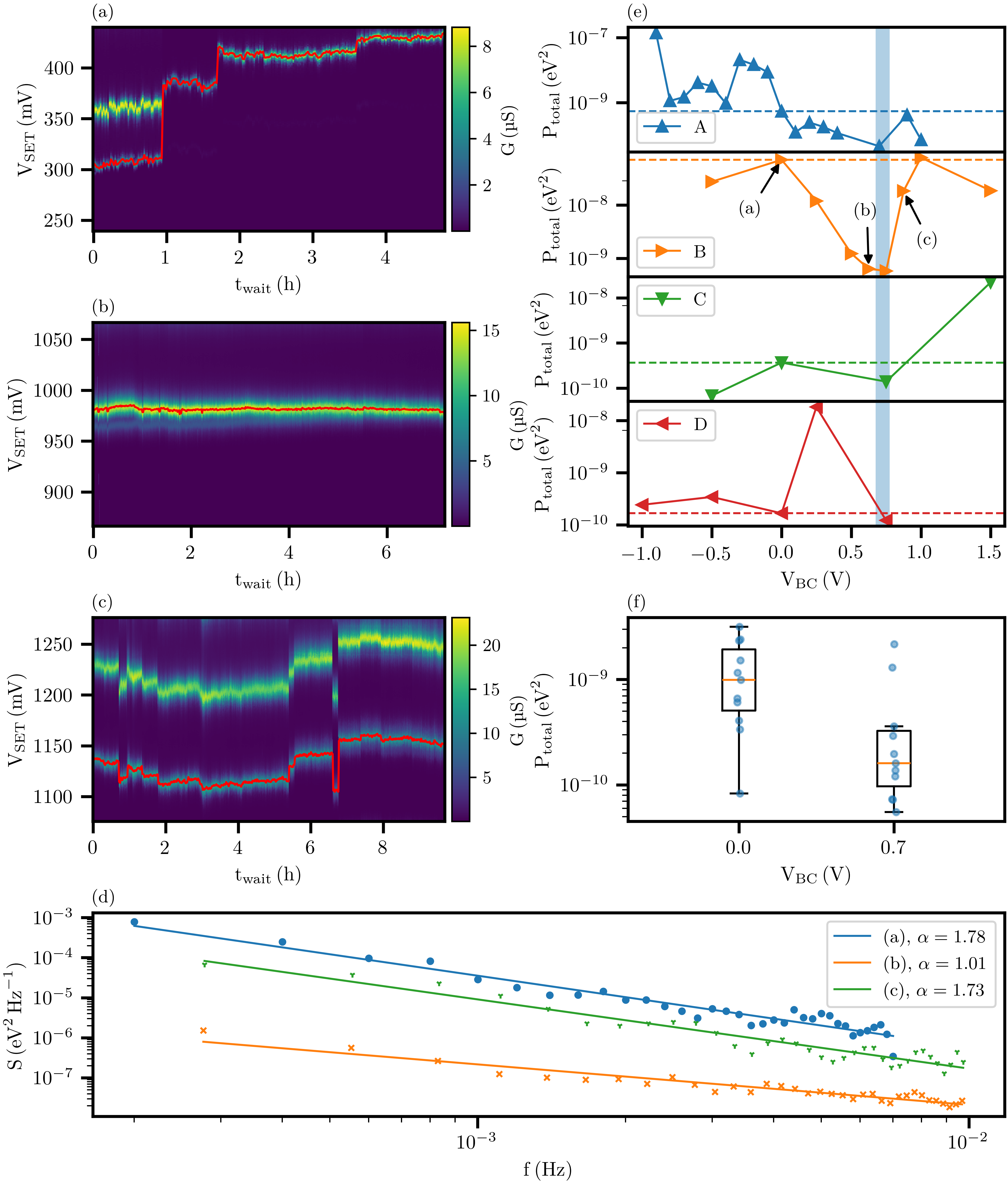}% Here is how to import EPS art
    \caption{\label{fig:nbcv}(a) - (c) Peaktracking measurements performed on Sample B for \SI{0}{\volt}, \SI{0.625}{\volt} and \SI{0.825}{\volt}. (b) Shows a vanishing side peak, which might indicate the presence of a parasitic dot. (d) Noise power spectral densities fitted with a $\beta/f^\alpha$ power spectrum. e) Integrated noise power versus bias cooling voltage. While the noise increases towards negative biases for most samples, a local minimum at \SI{0.7}{\volt} is found in each trace. The dashed line denotes the \SI{0}{\volt} noise level. (f) Boxplot of the integrated noise power of 22 peaktrackings performed in alternating \SI{0}{\volt}/\SI{0.7}{\volt} cooldowns. The box includes the interquartile (IQR) range and the whiskers extend up to 1.5 times the IQR.The orange solid line displays the median value. The median value of \SI{0.7}{\volt} is almost an order of magnitude lower than for \SI{0}{\volt}.}
\end{figure*}
The shown result was recorded on sample A after being cooled down with \SI{0}{\volt}.
We extract the exact peak position by fitting each individual trace with a secant function and define the position of its maximum as the peak position. The conversion of $\mathrm{V_{SET}}$ to \si{\electronvolt} is done by a static lever arm of 0.039\,eV/V. This value is the average of multiple extracted lever arms measured (using Coulomb diamond measurements \cite{Hanson2007}) for different samples and bias cooling voltages spanning a range from $0.035$ to 0.040\,eV/V.

Next, we perform a Welch-estimation of the power-spectral-density using the standard signal.welch method from the scipy python package. To exclude artificial frequencies arising from drift in between the start and the end point of a trace, we additionally employ a Hann-type windowing function. The estimation is always performed on the first measured peak. The result can be found in Fig. \ref{fig:metrics_xplain} (b). The resulting spectrum follows a $1/f^\alpha$-noise distribution with $\alpha = ~1.57$. This lies in between $\alpha \approx 1$, suggesting the presence of many two level fluctuators with a broad distribution of time scales \cite{Paladino2014}, and $\alpha \approx 2$, indicating the presence of random walk noise \cite{Barnes1966}. Following, we integrate the fitted spectra in the band of \SI{200}{\micro\hertz} to \SI{10}{\milli\hertz}, highlighted in purple in Fig. \ref{fig:metrics_xplain} (b). Quoting the total noise power instead of the e.g. \SI{1}{\milli\hertz}-noise has the advantage that all frequencies in the measured band contribute to the noise value. This metric is chosen because we want to weigh both, the rarely occurring large jumps as well as the small displacments in peak position that happen in between every single datapoint.

The procedure is applied to all measured peaktracks. A selection measured on sample B is found in Fig. \ref{fig:nbcv} (a) to (c). Peaktracks dominated by large jumps tend towards $\alpha \approx 2$, while peaktracks which are dominated by fluctuations around the original working point tend towards $\alpha \approx 1$. Resulting noise spectra are shown in Fig. \ref{fig:nbcv} (d). The noise power versus bias cooling voltage can be seen in Fig. \ref{fig:nbcv} (e). Four samples show a significant noise reduction at \SI{0,7}{\volt}. Sample B shows an improvement in the integrated noise power by two orders of magnitude. Towards more positive voltages, samples B and C show an increase in noise.

We identify two types of noise on the SET peak position. A high frequency, low amplitude fluctuation and rarely occurring jumps with a large amplitude. These jumps occur on the timescale of hours and drastically affect the noise performance (Fig. \ref{fig:nbcv}). Since the individual peak-tracking measurements have a length of five hours, they may not lead to statistically significant data. To gain insight into the significance of our observations, we performed a measurement campaign consisting of 22 cooldowns of sample A. We measured eleven times the zerobias followed by the \SI{0,7}{\volt} bias.
% \begin{figure}
% \includegraphics{figures/stat_boxplot.png}% Here is how to import EPS art
% \caption{\label{fig:boxplot} Boxplot of the integrated noise power of 22 peaktrackings. The orange solid line displays the median value. The median value of \SI{0.7}{\volt} is almost an order of magnitude lower than for \SI{0}{\volt}.}
% \end{figure}
The results of the noise power integrated from \SI{200}{\micro\hertz} to \SI{10}{\milli\hertz} are shown in the Fig. \ref{fig:nbcv} (f). The orange line shows the median. The median value of the $0.7$\,V measurements is reduced by a factor of 6 in comparison to the median of the \SI{0}{\volt} measurements. To determine the statistical significance of the measured results, we performed an unpaired t-test comparing the measured data sets for \SI{0}{\volt} and \SI{0.7}{\volt}, which results in a probability of \SI{0.143}{\percent} that the two measured sample means arise from the same normal distribution.

\begin{figure}
    \includegraphics{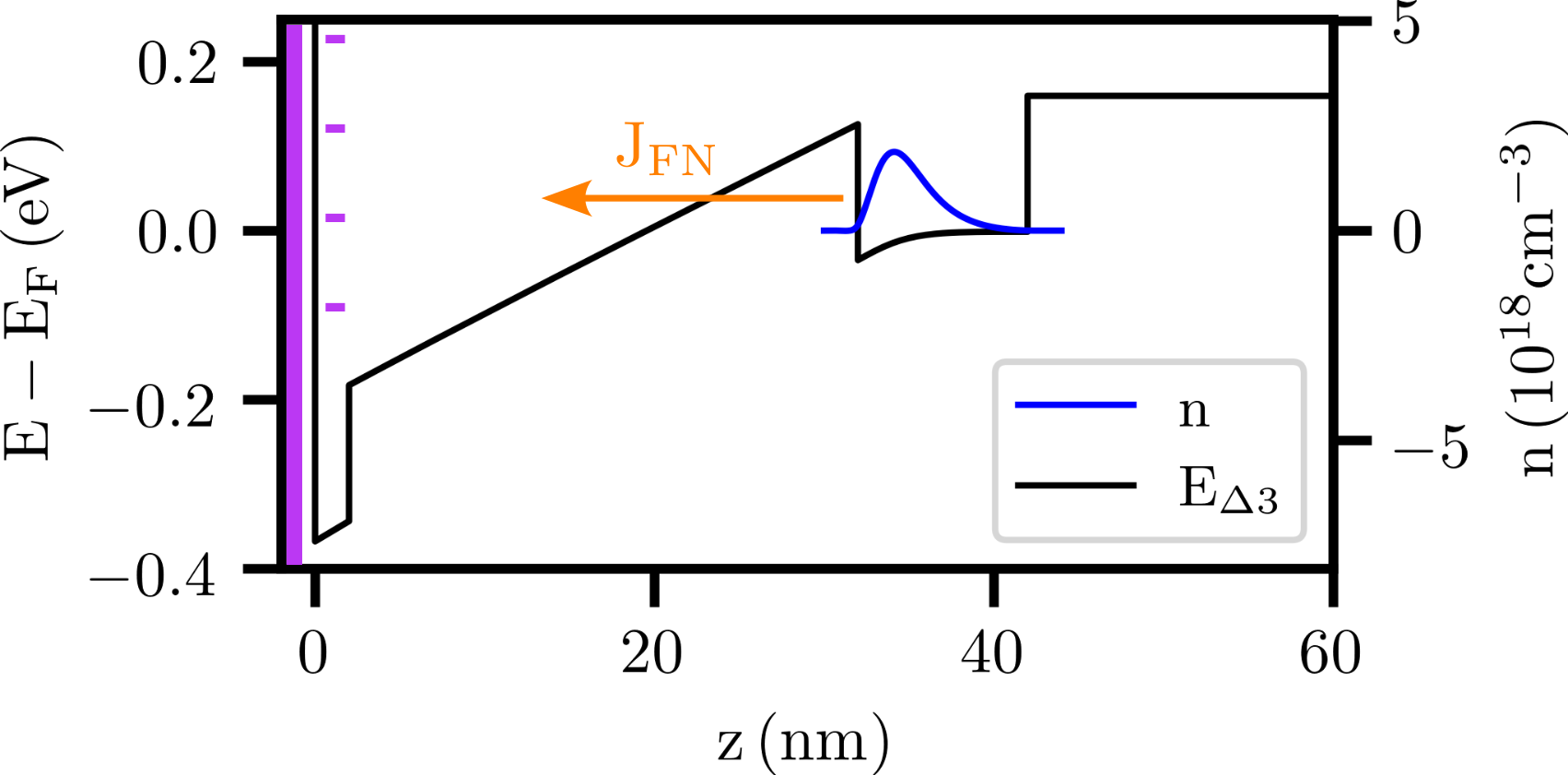}% Here is how to import EPS art
    \caption{\label{fig:fn_bandalignement}Simulated band structure alignment at the working point of a sample cooled down with $\mathrm{V_{BC} = \SI{0.7}{\volt}}$. $\mathrm{E_{\Delta3}}$ is the energy of the lowest conduction band. The violet region indicates the assumed position of the interface charges. They were distributed in a \SI{1}{\nano\meter} thick sheet located \SI{1.2}{\nano\meter} above the Si-cap-oxide interface. The orange arrow indicates the direction of the Fowler-Nordheim tunneling current density $\mathrm{J_{FN}}$. The electron density n is plotted in blue.}
\end{figure}

\begin{figure*}
    \includegraphics[width=\textwidth]{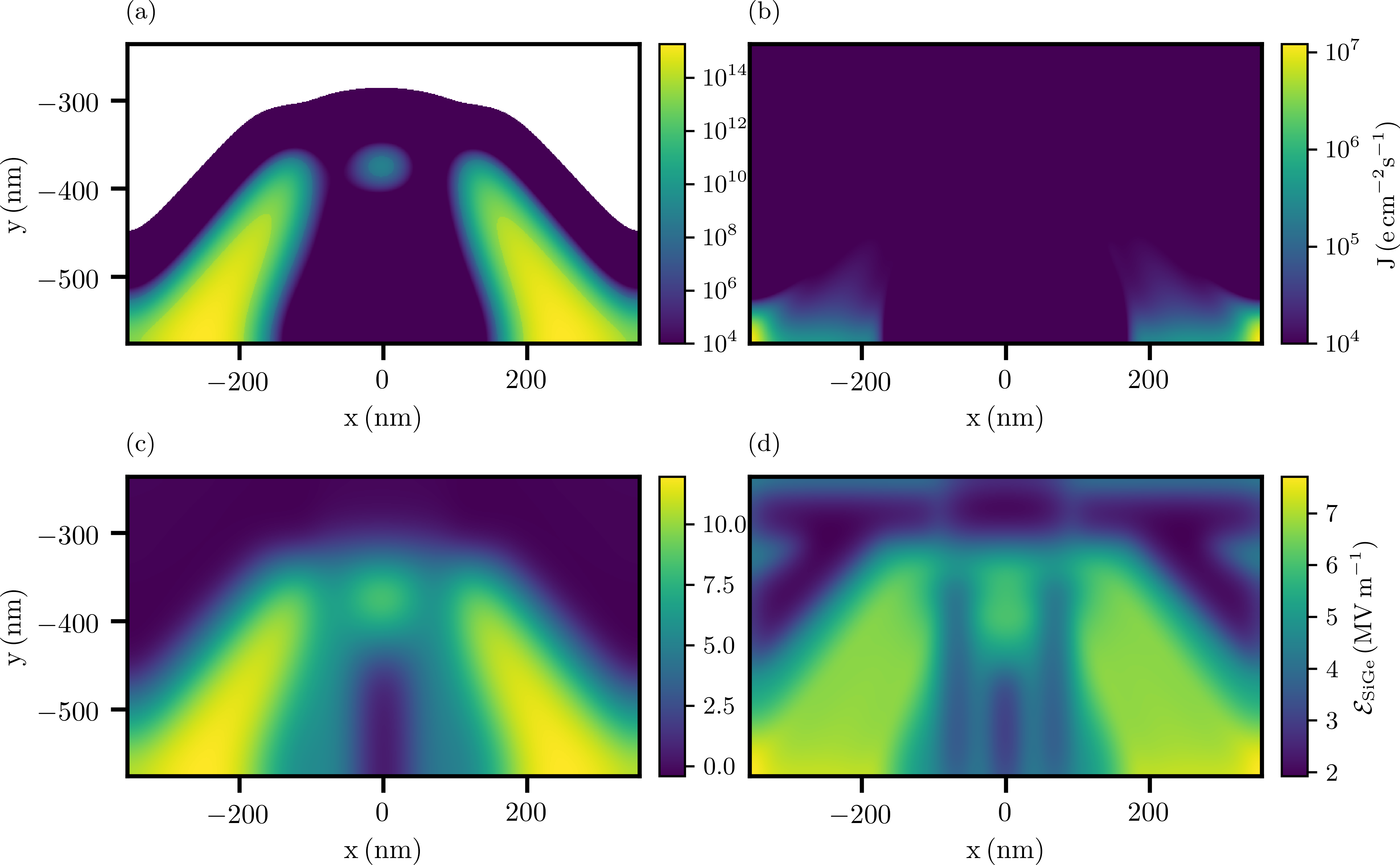}% Here is how to import EPS art
    \caption{\label{fig:tunnel_field}a) Simulated tunneling current from silicon channel into the silicon cap for a sample cooled with a \SI{0}{\volt} bias (a) and \SI{0.7}{\volt} bias (b). Simulated electric field distribution at the channel-spacer interface of a sample cooled with a \SI{0}{\volt} bias (c) and \SI{0.7}{\volt} bias (d).}
\end{figure*}

Three-dimensional Poisson-Schrödinger-simulations have been performed on the SET-area of the investigated devices. \ch{SiGe}-heterostacks are known to induce tunneling currents from the \ch{Si}-channel\cite{Huang2014} into the cap. Charge redistributions, as seen in Fig. \ref{fig:nbcv} (a), could be caused by local metal-to-insulator-transitions, triggered by tunneling into the cap \cite{Huang2014, Huang2015}. This interpretation is plausible, since the cap-oxide interface itself is MOSFET-like. In MOSFET devices the lengthscales of localization length versus potential variation \cite{Wilamowski2001} are known to create a percolation induced metal-to-insulator-transition \cite{Tracy2009}. In that case the 2DEG in the cap breaks down into charge-carrier-puddles containing mobile carriers. With the simulations we wish to verify that tunneling currents into the cap exist and can therefore source electrons which cause the charge redistributions. Bias cooling was included in simulation by placing interface-charges in the interstitial silicon-oxide layer under the metal gates. The interface-charge density under each gate-layer was calculated based on the turn-on voltage shift. We simulated the density needed to shift the turn-on voltage of gate layer $\mathrm{n}$ (with $\mathrm{n} \in \{1,2,3\}$) individually. The same charge density is assumed under each gate of a specific gate layer. To reflect the device's real working point, the gate-voltage configurations from individual measurements have been reproduced in simulation. The spatially large regions are treated (substrate, buffer, spacer) in Thomas-Fermi \cite{Thomas1927, Fermi1928, Tang2004} approximation. The regions where quantum confinement plays a major role (channel and cap),on the other hand, are treated with a self-consistent Schrödinger-Poisson approach in effective mass approximation \cite{Kane1959, Trellakis2006}. In Fig. \ref{fig:fn_bandalignement}, the simulated band diagram in growth direction for a sample cooled down with $\mathrm{V_{BC} = \SI{0.7}{\volt}}$ is shown. The barrier in between the silicon channel and the silicon cap-layer is of triangular shape, meaning that the tunneling currents can be calculated using the Fowler-Nordheim \cite{Fowler1928} model. With the resulting electric field and electrochemical potential, we calculate the tunneling rates from the SiGe-channel into the silicon cap. The spacially varying tunneling currents are shown in Fig. \ref{fig:tunnel_field} (a) for the \SI{0}{\volt} and (b) for the \SI{0.7}{\volt} bias cooling case. First, it is important to note that the maximum tunneling rate for the \SI{0.7}{\volt} case is reduced by seven orders of magnitude in comparison to the \SI{0}{\volt} case. Furthermore, the dot-region itself does not act as a source of electrons in the \SI{0.7}{\volt} case, meaning the number of tunneling electrons is not only reduced, but the tunneling events are mainly taking place further away from the region of interest. Generally, the bias cooled samples are operated at lower internal electric fields, as seen in Fig. \ref{fig:tunnel_field} (c) and (d).
In the bias-cooled-case the electric field difference in between the dot region and the barriers is larger. This means that a dot can be accumulated at lower accumulation gate fields, since the electric field of the barriers is more sharply defined due to the frozen-in charges, which are located at the Si-cap-\ch{Al2O3} interface.

\begin{figure}
\includegraphics{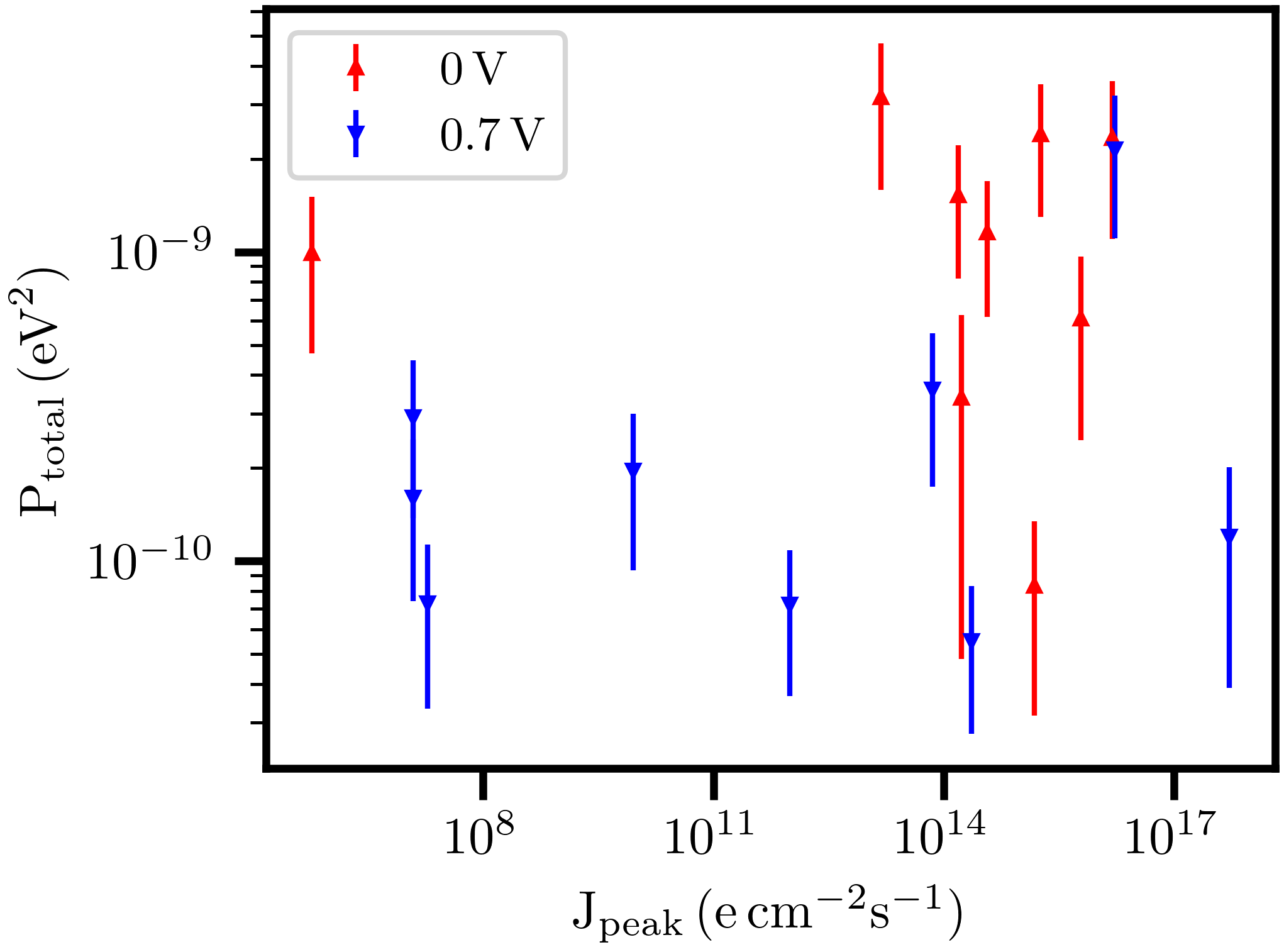}% Here is how to import EPS art
\caption{\label{fig:noisevstunnel} Measured integrated noise density versus simulated peak tunneling current. The datapoint belonging to the \SI{0.7}{\volt} dataset in the high-tunneling, high-noise quadrant shows an excess of telegraph noise which was only recorded once.}
\end{figure}

We repeated the simulation for each working point in Fig. \ref{fig:nbcv} (f), to verify an interrelation in between tunneling events and noise. In Fig. \ref{fig:noisevstunnel} the measured noise power is plotted versus the calculated maximum tunneling current. The \SI{0}{\volt} cases show a bunching towards high tunneling rates and high noise, whereas the \SI{0.7}{\volt} cases show bunching towards low tunneling and low noise. One possible explanation for the correlation of high noise and high tunneling rates is that the electrons flowing from the channel into the cap cause local metal-insulator transitions in the disordered silicon cap \cite{Huang2014}. The avalanche-like charge redistributions seen in the peaktracking measurements might be caused by a few excess electrons, locally exceeding the percolation density, triggering a large charge transfer, which is supported by the fact that the number of large charge redistributions is reduced in the \SI{0.7}{\volt} dataset. At high bias cooling voltage,  a high number of the acceptor defects located at the Si-cap \ch{Al2O3} interface are charged. We propose two main mechanisms to explain the increase in noise for voltages above \SI{0.7}{\volt}. First, the high density of negatively charged defects leads to working points with high electric fields. Additional tunneling could be a result. Furthermore, the increase in charged defects at the interface could lead to the individual defects exchanging charges, therefore effectively raising the noise again. For bias cooling voltages exceeding \SI{2}{\volt} the 2DEG is accumulated without the application of an accumulation voltage. Here, the trapped charges mimic the role of a dopant. For more negative voltages the noise is increasing in the case of sample A and D, not showing a global trend.

In conclusion, we found that bias cooling of undoped \ch{Si}/\ch{Si_x Ge_{1-x}} heterostacks causes charges to be trapped in between the silicon channel and the metal gates. This shifts the turn-on voltages linearly with the applied cooldown bias in the range of \SI{-1}{\volt} to \SI{1}{\volt}. We investigated the low frequency charge noise of bias cooled devices. For this we used peaktracking measurements. To quantify the noise for each bias cooling voltage we computed the total noise power for every peak track, which has a minimum around \SI{0.7}{\volt} bias cooling voltage. In samples A and B, a global minimum is visible in the total noise power. Samples C and D show a local minimum around \SI{0.7}{\volt}. In case of Sample B we could reduce the total noise power by a factor of 120 during the \SI{0.75}{\volt} cooldown in comparison to the \SI{0}{\volt} cooldown, and during a 22-cooldown campaign the measured median of the total noise power was reduced by a factor of 6 for the \SI{0.7}{\volt} cooldown in comparison to the \SI{0}{\volt} value. Our measurements show that a significant variation of the noise level can occur as a function of cooldown-bias and that this variation is not a random fluctuation from cooldown to cooldown. While a cooldown bias of \SI{0.7}{\volt} leads to good results on all four devices considered, more statistics would be needed to tell if this value reflects a device-independent optimum, and how the optimal cooldown strategy depends on the device design. Another important implication of the significant variation between cooldowns and bias voltages is that a lot of statistics is needed to draw reliable conclusions from noise measurements. In addition, we present the results of a three-dimensional Schrödinger-Poisson simulation, based on measured working points. Here, we find that samples cooled down with a \SI{0.7}{\volt} bias show a by seven orders of magnitude reduced tunneling current from the channel into the cap. While the direct proof remains elusive, we correlate the simulation to our noise measurement results, and find a bunching of datapoints in the high noise, high tunneling as well as in the low noise, low tunneling quadrants. As a next step, bias cooling could be extended to qubit samples, investigating the effect on coherence. Furthermore, tunneling in the cap has been proposed as one of the root causes of the quantum dot instabilities.

This work has been funded by the German Research Foundation (DFG) under Germany's Excellence Strategy - Cluster of Excellence Matter and Light for Quantum Computing" (ML4Q) EXC 2004/1 – 390534769 and the Gottfried Wilhelm Leibniz-Award, ZVN-2020\textunderscore WE 4458-5. Project Si-QuBus received funding from the QuantERA ERA-NET Cofund in Quantum Technologies implemented within the European Union's Horizon 2020 Programme. The device fabrication has been done at HNF - Helmholtz Nano Facility, Research Center Juelich GmbH \cite{albrecht_hnf_2017}. Furthermore the authors wish to thank Malte Neul for the organization of our meetings and for helpful discussions.
%TC:endignore
%TC:ignore
%\bibliography{citations.bib}% Produces the bibliography via BibTeX.

%TC:endignore
\end{document}